\documentstyle[12pt]{article}

\newcommand{\be}{\begin{equation}}
\newcommand{\ee}{\end{equation}}

\newcommand{\bea}{\begin{eqnarray}}
\newcommand{\eea}{\end{eqnarray}}

\newcommand{\p}{\partial}

\newcommand{\nn}{\nonumber \\}
\newcommand{\f}{\frac}
\newcommand{\w}{\wedge}

\begin{document}

\thispagestyle{empty}
\renewcommand{\thefootnote}{\fnsymbol{footnote}}

{\hfill \parbox{4cm}{ hep-th/0804.1994 \\
  }}

\bigskip\bigskip

\begin{center} \noindent \Large \bf Supersymmetric and
nonsupersymmetric perturbations to KT
\end{center}

\bigskip\bigskip\bigskip

\centerline{ \normalsize \bf Shesansu Sekhar Pal}

\bigskip
\bigskip\bigskip

\centerline{ \it Barchana,  Jajpur 754081, Orissa, India }
\centerline{ \textsf{shesansu {\frame{\shortstack{AT}}} gmail.com }}

\bigskip

\bigskip\bigskip

\renewcommand{\thefootnote}{\arabic{footnote}}

\centerline{\bf \small Abstract}

We studied the supersymmetric and non-supersymmetric perturbations
to cascading gauge theory. In particular we use KT background and
the back reaction of the generic  linearized perturbation make the
dilaton to run and the $T^{(1,1)}$ gets squashed which in turn make
the supersymmetry to be broken. But if we make a special linearized
perturbation in such way that the $T^{(1,1)}$ is not squashed then
the corresponding perturbation preserve supersymmetry.
\newpage
\medskip\section{Introduction}
It has been a practice to generate as much gravity solution as
possible. As these solutions are of prime importance because of the
celebrated gauge gravity duality \cite{jm}, \cite{gkp1}, \cite{ew}.
Even though this duality is for the  $AdS_5 \times S^5$ and the
${\cal N}=4$ in 3+1 dimension, but still there is a hope that we can
apply this duality for different cases  in the sense of
non-conformal, supersymmetric and non-supersymmetric solutions.

In this context there arises an interesting gravity solution
generated by Klebanov and collaborators, (Nekrasov, Tseytlin and
Strassler) \cite{kn},\cite{kt},\cite{ks}, who considered the
supersymmetric intersections of N D3 branes and M D5 branes wrapped
on the 2-cycle of a Calabi-Yau manifold that is the conifold
\cite{cd}. For this configuration they successfully  generated an
interesting and important gravity solutions which preserves $ {\cal
N}=1$ supersymmetry and are non-conformal in nature. The most
interesting solution that is \cite{ks} shows confinement but in the
large $r$ limit it goes over to the singular \cite{kt} solution. The
corresponding dual field theory is the $ {\cal N}=1$ supersymmetric
$SU(N+M)\times SU(N)$ gauge theory with two bi-fundamental and two
anti-bifundamental chiral superfields and with a non-trivial
superpotential. A T-dual version of this is studied by
\cite{au},\cite{dm}, in which the authors took the intersection of
two stacks of separated NS 5 branes extended along (12345) and
(12389) and a stack of D4 branes along (1236). The coordinate $x^6$
is taken as compact and the two gauge couplings are determined by
the position of the NS 5 brane, in particular the couplings are
equal when they are placed at the diametrically opposite points.

 These solutions has the interesting feature like  the
cascading behavior \cite{ns}, which means the rank of  the gauge
group falls and become $SU(N-M)\times SU(M)$ gauge theory after
going through one Seiberg duality and this process continues until
IR where it reaches a confining gauge theory described by SU(M)
gauge theory. The duality is properly constructed from the field
theory point of view in  \cite{ks}, \cite{ms} and its being
emphasized in \cite{ms} that this duality is not a property at IR
but its an exact duality.

The gauge theory has got two couplings $g_1,~~g_2$ apart from the
coupling $\lambda$ that comes from the superpotential. These two
gauge couplings are related to the dilaton and the integrated 2-from
flux that comes from NS-NS sector over the 2-cycle of $T^{(1,1)}$ by
gauge/gravity duality, whose precise form is given in \cite{ks},
\cite{ms}. Topologically $T^{(1,1)}$ is $S^2\times S^3$ with
symmetry $SU(2)\times SU(2)\times U(1)$, and  is a coset space
$\frac{SU(2)\times SU(2)}{U(1)}$.

From the gravity point of view the solutions \cite{kt}, \cite{ks}
has got a non-trivial 3-from complex field strength, $G_3$, which
has the structure (2,1),   according to its Hodge classification
\cite{ks}, \cite{gkp} and  obey the ISD condition. This particular
structure of the flux  in turn make the dilaton to be a constant and
the brane configuration supersymmetric. But its not necessarily true
that by making the dilaton to run we have to break supersymmetry. We
shall see both the  supersymmetric and non-supersymmetric
(linearized) solution for which  the dilaton is not constant.

The metric of the KT solution  preserves  a U(1) symmetry, which is
associated  to the shift of an angle connected to the fiber of
$T^{(1,1)}$, but is broken to $Z_{2M}$ by the 2-form potential
coming from the RR sector, even though its field strength does
preserve this  symmetry. This in the dual field theory is
interpreted as the break down of the R-symmetry \cite{kow} and its a
spontaneous breaking.  From this it follows that in the absence of
D5 branes wrapped on the 2-cycle this U(1) symmetry is an exact
symmetry. In the linearized perturbed solution, we shall see
explicitly that the whole things goes over and the R-symmetry is
broken irrespective of whether  we break supersymmetry or not, as we
do not change the 2-form RR potential, $C_2$.

Summarizing, in this paper, we shall see the following points. \\
(1) By doing a linearized perturbation to the metric and fluxes with
two independent parameters ${\cal S}$ and $\phi$, \cite{dkm}, we
considered the back reaction of this perturbation to the original KT
solution and obtained the solution in Einstein frame as
\bea\label{solution}
ds^2&=&h^{-\frac{1}{2}}\eta_{\mu\nu}dx^{\mu}dx^{\nu}+
h^{\frac{1}{2}}[dr^2+r^2(1+\frac{16}{81}\frac{h_2 {\cal
S}}{r^4})\frac{g^2_5}{9}+\frac{r^2}{6} (g^2_1+g^2_2+g^2_3+g^2_4)]\nn
F_3&=& \frac{M \alpha'}{4} g_5\w(g_1\w g_2+g_3\w g_4),~~C_0=0\nn
B_2&=&\frac{g_s M \alpha'}{2}f(r) (g_1\w g_2+g_3\w
g_4),~H_3=\frac{g_s M \alpha'}{2}f'(r) dr\w (g_1\w g_2+g_3\w g_4)\nn
\widetilde{F_5}&=&(1+\star_{10}){\cal F}_5,~~ {\cal F}_5=\frac{g_s
M^2 \alpha'^2}{4} \ell(r) g_1\w g_2\w g_3\w g_4\w g_5\nn
h(r)&=&\frac{27 \pi g_s \alpha'^2}{4r^4}[\frac{3g_s M^2}{2\pi} Log
~\frac{r}{r_0}+\frac{3g_s M^2}{8\pi}]+\f{(g_S M
\alpha')^2}{r^8}[(h_1+h_2 Log~r){\cal S}-h_3
\phi]\nn\Phi(r)&=&Log~g_s+\f{1}{r^4}\bigg((-\frac{64}{81}
h_1+\frac{52}{81}h_2){\cal S}-\frac{16}{27}h_2 {\cal S}
Log~r+\frac{64}{81}h_3 \phi\bigg)\nn  f(r)&=&k(r)=\ell(r)=\f{3}{2}
Log~r+\f{1}{r^4}\bigg((\frac{8}{27}h_1-\frac{h_2}{27}){\cal
S}+\frac{4}{9}{\cal S}h_2 Log~r-\frac{8}{27}h_3 \phi\bigg),\nn\eea
where $h_1,~h_2$ and $h_3$ are real and independent parameters. For
a very specific choice like: $h_1=\f{1053}{256},~~h_2=\f{81}{16}$,
and $h_3=\f{81}{64}$ , we get the solution \cite{dkm}.

(2)From this solution it just follows that we have not changed the
RR 3-form, $F_3$, flux and hence the quantization condition
associated to it remains intact. But the $H_3$ is now different so
also the 5-form flux. The corresponding quantization condition has
the leading term which goes logarithmically but the subleading term
goes as inverse power law for $h_2=0$ and in general  \be
\oint_{T^{(1,1)}} \widetilde{F_5} \sim g_s M^2 \alpha'^2\bigg[
\f{3}{2} Log~r+\f{1}{r^4}\bigg((\frac{8}{27}h_1-\frac{h_2}{27}){\cal
S}+\frac{4}{9}{\cal S}h_2 Log~r-\frac{8}{27}h_3 \phi\bigg)\bigg] \ee

(3) The complex 3-from flux made out of NS-NS, RR 3-from fluxes and
the dilaton contains both the imaginary self dual and the imaginary
anti self dual piece. \bea G_+&=&\frac{i M
\alpha'}{r^9}\bigg[2-\frac{{\cal S}
h_2}{r^4}\bigg(\frac{28}{81}+\frac{16}{27}
Log~r\bigg)\bigg]\overline{z}_m dz_m\w\epsilon_{ijkl}z_i
\overline{z}_jdz_k\w \overline{dz}_l\nn G_-&=&-\frac{i M
\alpha'}{r^9}\bigg[\frac{{\cal S}
h_2}{r^4}\bigg(\frac{12}{81}+\frac{16}{27} Log~r\bigg)\bigg]z_m
\overline{dz}_m\w\epsilon_{ijkl}z_i \overline{z}_jdz_k\w
\overline{dz}_l.\eea It  means generically the solution presented in
eq(\ref{solution}) break supersymmetry. However for a very specific
choice of parameter the solution preserves supersymmetry and it is
$h_2=0$. Note that the squashing of $T^{(1,1)}$ is  proportional to
$h_2$.

(4) Upon looking at the space of solutions, it just follows that
there is a supersymmetry preserving 2-plane described by $(h_1,h_3)$
which sits at $h_2=0$. Away from this particular plane supersymmetry
is broken and is broken dynamically.

(5)From the solution presented in eq(\ref{solution}), it just
follows trivially that the dilaton runs irrespective of whether the
solution is supersymmetric or not. But the way it runs depends very
much on whether supersymmetry is preserved or not.

 The organization of the paper is as follows. In section 2,
we shall write down the equations of motion for IIB and the ansatz
that we are going to use and in section 3, we shall give the details
of the solutions and then conclude in section 4.

\section{Ansatz and the equations}
We would like to set up the notation for which we can apply the
resulting equations for the deformed conifold. Let us introduce some
1-form objects following \cite{mt},\cite{ks}\bea
&&g_1=\f{e_1-e_3}{\sqrt{2}},~~~~ g_2=\f{e_2-e_4}{\sqrt{2}}\nn
&&g_3=\f{e_1+e_3}{\sqrt{2}},~~~~
g_4=\f{e_2+e_4}{\sqrt{2}},~~~~g_5=e_5, \eea where \bea
&&e_1=-sin\theta_1 d\phi_1,~~~~e_2=d\theta_1,~~~~~e_3=cos\psi
~sin\theta_2 d\phi_2-sin\psi d\theta_2 \nn &&e_4=sin\psi~sin\theta_2
d\phi_2+cos\psi d\theta_2,e_5=d\psi+cos\theta_1 d\phi_1+cos\theta_2
d\phi_2. \eea and construct an ansatz of  10-dim metric consistent
with the symmetry  of a deformed conifold

\be\label{metric_ansatz}
ds^2=A^2(\tau)\eta_{\mu\nu}dx^{\mu}dx^{\nu}+B^2(\tau)
d\tau^2+C^2(\tau)g^2_5+D^2(\tau)(g^2_3+g^2_4)+E^2(\tau)(g^2_1+g^2_2).\ee

The 3-form fields, dilaton, axion and 5-form self-dual field is
assumed to take the following form \bea
F_3&=&\f{M\alpha'}{2}[(1-F)g_5\w g_3\w g_4+F g_5\w g_1\w g_2+F'
d\tau\w (g_1\w g_3+g_2\w g_4)]\nn B_2&=&\f{g_s M \alpha'}{2}[f(\tau)
g_1\w g_2 +k(\tau) g_3\w g_4],~~C_0=0,~~ \Phi=\Phi(\tau)\nn
H_3&=&\f{g_s M \alpha'}{2}[d\tau\w(f' g_1\w g_2+k' g_3\w
g_4)+\f{k-f}{2}g_5\w(g_1\w g_3+g_2\w g_4)]\nn {\tilde
F}_5&=&(1+\star_{10}){\cal F}_5,~~{\cal F}_5=\f{g_s M^2
\alpha'^{2}}{4} {\ell} g_1\w g_2\w g_3\w g_4\w g_5, ~~{\ell
}=f(1-F)+k F\nn \eea

Note that the  $H_3$ we have taken is different from the ansatz
considered in \cite{pt}. The difference is that our ansatz has a
symmetry (ant-symmetric)  under $Z_2$, which is to interchange the
two $S^2$'s i.e. $(\theta_1,\phi_1)$ with $(\theta_2,\phi_2)$. Its
easy to check that the equation of motion remain unchanged under
this.

For completeness we shall write down the ansatz to NS-NS 3-form
field strength, written down in \cite{pt} \bea
H^{(PT)}_3&=&(h'_2-h'_1)d\tau\w g_1\w g_2-(h'_1+h'_2) d\tau\w g_3\w
g_4-\nn&& h_2 g_5\w(g_1\w g_3+g_2\w g_4)+2 \chi' d\tau\w(g_2\w
g_3-g_1\w g_4),\eea where we have taken $\tau$ as the radial
coordinate.

This form of $H^{(PT)}_3$ has a piece which is symmetric and another
piece anti-symmetric under $Z_2$, which means if we want to have the
$Z_2$ symmetry  in the equation of motion then we need to set
$\chi'$ to zero. In any case we shall take the ansatz written above.

 The dilaton equation in this case is \be \nabla^2 \Phi=-\f{g_s
e^{-\Phi}}{12}[H_{MNP}H^{MNP}-e^{2\Phi}F_{MNP}F^{MNP}].\ee

The various terms in this equation are \bea \nabla^2
\Phi&=&\f{1}{A^4BCD^2E^2 }\p_{\tau}[\f{A^4 C D^2 E^2 \Phi'}{B}]\nn
H_{MNP}H^{MNP}&\pm&
e^{2\Phi}F_{MNP}F^{MNP}=\f{3}{2}M^2\alpha'^{2}[\f{g^2_s f'^{2}}{B^2
E^4}+\f{g^2_s k'^{2}}{B^2 D^4}+\f{(k-f)^2}{2}\f{g^2_s}{C^2 D^2
E^2}\nn && \pm
 e^{2\Phi}\f{(1-F)^2}{C^2 D^4}\pm e^{2\Phi}\f{F^2}{C^2 E^4}\pm
 2e^{2\Phi}\f{F'^{2}}{B^2 D^2 E^2}]\eea So the dilaton equation of
motion give \bea\label{dilaton_eq}&& \f{1}{A^4BCD^2E^2
}\p_{\tau}[\f{A^4 C D^2 E^2 \Phi'}{B}]=-\f{g_s M^2\alpha'^{2}}{8}
e^{-\Phi}[\f{g^2_s f'^{2}}{B^2 E^4}+\f{g^2_s k'^{2}}{B^2
D^4}+\nn&&\f{(k-f)^2}{2}\f{g^2_s}{C^2 D^2 E^2}-
e^{2\Phi}\f{(1-F)^2}{C^2 D^4}-e^{2\Phi}\f{F^2}{C^2
E^4}-2e^{2\Phi}\f{F'^{2}}{B^2 D^2 E^2}]\eea

The equation for $F_3$ is \be d\star_{10} (e^{\Phi} F_3)=g_s{\tilde
F}_5\w H_3.\ee The expression for both the LHS and RHS are

\bea d\star_{10} (e^{\Phi} F_3)&=&\f{M\alpha'}{2}[\f{1-F}{2}
e^{\Phi}\f{A^4 B E^2}{CD^2}-\f{F}{2}
e^{\Phi}\f{A^4BD^2}{CE^2}+\p_\tau(F' e^{\Phi} \f{A^4C}{B})]\nn &
&d\tau\w g_5\w(g_1\w g_3+g_2\w g_4)\w dx^0\cdots\w dx^3\nn
g_s{\tilde F}_5\w H_3&=&\f{(g_s M\alpha')^3}{8}
\f{A^4B}{CD^2E^2}\f{\ell(k-f)}{2}\nn&&d\tau\w g_5\w(g_1\w g_3+g_2\w
g_4)\w dx^0\cdots\w dx^3 \nn\eea

Hence the resulting equation of motion for $F_3$ is
\be\label{F_3_eq} \f{1-F}{2} e^{\Phi}\f{A^4 B E^2}{CD^2}-\f{F}{2}
e^{\Phi}\f{A^4BD^2}{CE^2}+\p_\tau(F' e^{\Phi} \f{A^4C}{B})= \f{g^3_s
M^2\alpha'^2}{4} \f{A^4B}{CD^2E^2}\f{\ell(k-f)}{2}.\ee

The equation for $H_3$ is \be d\star_{10} (e^{-\Phi}
H_3)=-g_s{\tilde F}_5\w F_3.\ee The expression for both the LHS and
RHS are

\bea d\star_{10} (e^{-\Phi} H_3)&=&-\f{g_s M \alpha'}{2}[\p_{\tau}
(f' e^{-\Phi}\f{A^4 CD^2}{BE^2})+\f{k-f}{2} e^{-\Phi}\f{A^4
B}{C}]g_5\w g_3\w g_4\w dx^0\cdots\w dx^3\nn&-&\f{g_s M
\alpha'}{2}[\p_{\tau} (k' e^{-\Phi}\f{A^4 CE^2}{BD^2})-\f{k-f}{2}
e^{-\Phi}\f{A^4 B}{C}]g_5\w g_1\w g_2\w dx^0\cdots\w dx^3\nn
-g_s{\tilde F}_5\w F_3&=&\f{g^2_s M^3 \alpha'^3}{8} \f{A^4 B}{CD^2
E^2}\ell dx^0\cdots\w dx^3\w d\tau\w[(1-F)g_5\w g_3\w g_4\nn
&&+Fg_5\w g_1\w g_2]\eea

Hence the resulting equation of motion for $H_3$ are
\bea\label{H_3_eq} \f{g_s M^2 \alpha'^2}{2}\f{A^4 B}{CD^2 E^2}\ell
(1-F)&=&2 \p_{\tau} (f' e^{-\Phi}\f{A^4 CD^2}{BE^2})+(k-f)
e^{-\Phi}\f{A^4 B}{C}\nn \f{g_s M^2 \alpha'^2}{2}\f{A^4 B}{CD^2
E^2}F \ell&=&2 \p_{\tau} (k' e^{-\Phi}\f{A^4 CE^2}{BD^2})-(k-f)
e^{-\Phi}\f{A^4 B}{C}.\eea

The Bianchi identities associated to the form fields are identically
satisfied.

Now the equation of motion for the metric results \bea
R_{MN}&=&\f{1}{2}\p_M \Phi \p_N \Phi+\f{g^2_s}{96} {\tilde
F}_{MPQRS}{{\tilde F}_N}{}^{PQRS}+ \f{g_s
e^{-\Phi}}{4}[H_{MPQ}{H_N}^{PQ}+e^{2\Phi}F_{MPQ}{F_N}^{PQ}]\nn
&-&\f{g_s
e^{-\Phi}}{48}g_{MN}[H_{PQR}H^{PQR}+e^{2\Phi}F_{PQR}F^{PQR}].\eea

For the choice of our metric and form fields we get the following
components of  Ricci tensor \bea
R_{\mu\nu}&=&-\eta_{\mu\nu}\bigg[\f{3A'^2}{B^2}-\f{AA'B'}{B^3}+\f{AA'C'}{B^2C}+\f{2AA'D'}{B^2D}+\f{2AA'E'}{B^2E}+\f{AA''}{B^2}\bigg]\nn
&=&-\eta_{\mu\nu}\bigg[\f{(g_s M\alpha')^4}{64}\f{\ell^2 A^2}{C^2
D^4 E^4}+\f{g_s M^2 \alpha'^2}{32}A^2 e^{-\Phi}\bigg(\f{g^2_s
f'^{2}}{B^2 E^4}+\f{g^2_s k'^{2}}{B^2 D^4}+\nn &
&\f{(k-f)^2}{2}\f{g^2_s}{C^2 D^2 E^2}+
 e^{2\Phi}\f{(1-F)^2}{C^2 D^4}+ e^{2\Phi}\f{F^2}{C^2 E^4}+
 2e^{2\Phi}\f{F'^{2}}{B^2 D^2 E^2}\bigg)\bigg].\nn\eea

\bea
R_{\tau\tau}&=&\Bigg(\f{4A'B'}{AB}+\f{B'C'}{BC}+\f{2B'D'}{BD}+\f{2B'E'}{BE}-\f{4A''}{A}-\f{C''}{C}-\f{2D''}{D}-\f{2E''}{E}
\Bigg)\nn&=&\f{1}{2}\Phi'^2-\f{(g_s M\alpha')^4}{64}\f{\ell^2
B^2}{C^2 D^4 E^4}+ \f{g_s M^2 \alpha'^2}{8} e^{-\Phi}\bigg(\f{g^2_s
f'^{2}}{ E^4}+\f{g^2_s k'^{2}}{ D^4}+ 2 e^{2\Phi}\f{F'^2}{ D^2
E^2}\bigg)-\nn &&\f{g_s M^2 \alpha'^2}{32} e^{-\Phi}\bigg(\f{g^2_s
f'^{2}}{ E^4}+\f{g^2_s k'^{2}}{ D^4}+ \f{(k-f)^2}{2}\f{g^2_s
B^2}{C^2 D^2 E^2}+
 e^{2\Phi}\f{(1-F)^2B^2}{C^2 D^4}+\nn && e^{2\Phi}\f{F^2B^2}{C^2 E^4}+
 2e^{2\Phi}\f{F'^{2}}{ D^2 E^2}\bigg)\eea

\bea
R_{\theta_1\theta_1}&=&R_{\theta_2\theta_2}=1-\f{C^2}{4D^2}-\f{D^2}{16C^2}-\f{C^2}{4E^2}+\f{D^4}{16C^2E^2}-\f{E^2}{16C^2}+\f{E^4}{16C^2D^2}-\f{2DA'D'}{AB^2}
+\nn&&\f{DB'D'}{2B^3}-\f{DC'D'}{2B^2C}-\f{D'^2}{2B^2}-\f{2EA'E'}{AB^2}+\f{EB'E'}{2B^3}-\f{EC'E'}{2B^2C}-\f{DD'E'}{B^2E}-\nn&&\f{ED'E'}{B^2D}-
\f{E'^2}{2B^2}-\f{DD''}{2B^2}-\f{EE''}{2B^2} \nn&=&\f{(g_s
M\alpha')^4}{128}\f{\ell^2 (D^2+E^2)}{C^2 D^4 E^4}+\f{g_s M^2
\alpha'^2}{16}e^{-\Phi}\bigg(\f{g^2_s f'^{2}}{B^2 E^2}+\f{g^2_s
k'^{2}}{ B^2D^2}+\f{g^2_s (k-f)^{2}}{4C^2 D^2}+\nn&&\f{g^2_s
(k-f)^{2}}{4C^2 E^2}+e^{2\Phi}\f{(1-F)^2}{ C^2
D^2}+e^{2\Phi}\f{F^2}{ C^2 E^2}+e^{2\Phi}\f{F'^2}{ B^2 D^2}+
e^{2\Phi}\f{F'^2}{ B^2 E^2}\bigg)-\f{g_s M^2
\alpha'^2}{64}\nn&&e^{-\Phi}(D^2+E^2)\bigg(\f{g^2_s f'^{2}}{B^2
E^4}+\f{g^2_s k'^{2}}{B^2 D^4}+\f{(k-f)^2}{2}\f{g^2_s}{C^2 D^2 E^2}+
 e^{2\Phi}\f{(1-F)^2}{C^2 D^4}\nn&&+ e^{2\Phi}\f{F^2}{C^2 E^4}+
 2e^{2\Phi}\f{F'^{2}}{B^2 D^2 E^2}\bigg).\eea

\be R_{\phi_i\phi_i}=sin^2\theta_i
R_{\theta_i\theta_i}+cos^2\theta_i R_{\psi\psi}\ee

\bea
R_{\psi\psi}&=&\bigg[\f{1}{2}+\f{C^4}{D^2E^2}-\f{D^2}{4E^2}-\f{E^2}{4D^2}-\f{4CA'C'}{AB^2}+\f{CB'C'}{B^3}-\f{2CC'D'}{B^2D}-
\f{2CC'E'}{B^2E}-\f{CC''}{B^2}\bigg]\nn &=&\f{(g_s
M\alpha')^4}{64}\f{\ell^2 }{ D^4 E^4}+\f{g_s M^2
\alpha'^2}{16}e^{-\Phi}\bigg(\f{g^2_s (k-f)^{2}}{
D^2E^2}+e^{2\Phi}\f{2(1-F)^2}{  D^4}+e^{2\Phi}\f{2F^2}{
E^4}\bigg)-\nn&&\f{g_s M^2 \alpha'^2}{32}e^{-\Phi} C^2\bigg(\f{g^2_s
f'^{2}}{B^2 E^4}+\f{g^2_s k'^{2}}{B^2
D^4}+\f{(k-f)^2}{2}\f{g^2_s}{C^2 D^2 E^2}+
 e^{2\Phi}\f{(1-F)^2}{C^2 D^4}\nn&&+ e^{2\Phi}\f{F^2}{C^2 E^4}+
 2e^{2\Phi}\f{F'^{2}}{B^2 D^2 E^2}\bigg).\eea

\bea
R_{\psi\phi_i}&=&cos~\theta_i\bigg[\f{1}{2}+\f{C^4}{D^2E^2}-\f{D^2}{4E^2}-\f{E^2}{4D^2}-\f{4CA'C'}{AB^2}+\f{CB'C'}{B^3}\nn&&-\f{2CC'D'}{B^2D}-
\f{2CC'E'}{B^2E}-\f{CC''}{B^2}\bigg]\nn&=&cos~\theta_i\bigg[\f{(g_s
M\alpha')^4}{64}\f{\ell^2 }{ D^4 E^4}+\f{g_s M^2
\alpha'^2}{16}e^{-\Phi}\bigg(\f{g^2_s (k-f)^{2}}{
D^2E^2}+e^{2\Phi}\f{2(1-F)^2}{  D^4}+e^{2\Phi}\f{2F^2}{
E^4}\bigg)-\nn&&\f{g_s M^2 \alpha'^2}{32}e^{-\Phi} C^2\bigg(\f{g^2_s
f'^{2}}{B^2 E^4}+\f{g^2_s k'^{2}}{B^2
D^4}+\f{(k-f)^2}{2}\f{g^2_s}{C^2 D^2 E^2}+
 e^{2\Phi}\f{(1-F)^2}{C^2 D^4}\nn&&+ e^{2\Phi}\f{F^2}{C^2 E^4}+
 2e^{2\Phi}\f{F'^{2}}{B^2 D^2 E^2}\bigg)\bigg]=cos~\theta_iR_{\psi\psi}.\eea

\bea
R_{\phi_1\phi_2}&=&-sin\theta_1sin\theta_2cos\psi\bigg[\f{C^2}{4D^2}+\f{D^2}{16C^2}-\f{C^2}{4E^2}+\f{D^4}{16C^2E^2}-\f{E^2}{16C^2}-
\f{E^4}{16C^2D^2}-\nn&&
\f{2DA'D'}{AB^2}+\f{DB'D'}{2B^3}-\f{DC'D'}{2B^2C}-\f{D'^2}{2B^2}+\f{2EA'E'}{AB^2}-\f{EB'E'}{2B^3}+
\f{EC'E'}{2B^2C}-\nn&&\f{DD'E'}{B^2E}
+\f{ED'E'}{B^2D}+\f{E'^2}{2B^2}-\f{DD''}{2B^2}+\f{EE''}{2B^2}
\bigg]+cos~\theta_1cos\theta_2\bigg[\f{1}{2}+\f{C^4}{D^2E^2}-\nn&&\f{D^2}{4E^2}-\f{E^2}{4D^2}-\f{4CA'C'}{AB^2}+\f{CB'C'}{B^3}-
\f{2CC'D'}{B^2D}-
\f{2CC'E'}{B^2E}-\f{CC''}{B^2}\bigg]\nn&=&\f{1}{2}sin\theta_1sin\theta_2cos\psi\bigg[\f{(g_s
M\alpha')^4}{64}\f{\ell^2 (E^2-D^2)}{C^2 D^4 E^4}+\f{g_s M^2
\alpha'^2}{8}e^{-\Phi}\bigg(\f{g^2_s f'^{2}}{B^2 E^2}-\f{g^2_s
k'^{2}}{ B^2D^2}+\nn&&\f{g^2_s (k-f)^{2}}{4C^2 D^2}-\f{g^2_s
(k-f)^{2}}{4C^2 E^2}-e^{2\Phi}\f{(1-F)^2}{ C^2
D^2}+e^{2\Phi}\f{F^2}{ C^2 E^2}+e^{2\Phi}\f{F'^2}{ B^2 D^2}-
e^{2\Phi}\f{F'^2}{ B^2 E^2}\bigg)\nn&&-\f{g_s M^2
\alpha'^2}{32}e^{-\Phi}(E^2-D^2)\bigg(\f{g^2_s f'^{2}}{B^2
E^4}+\f{g^2_s k'^{2}}{B^2 D^4}+\f{(k-f)^2}{2}\f{g^2_s}{C^2 D^2 E^2}+
 e^{2\Phi}\f{(1-F)^2}{C^2 D^4}\nn&&+ e^{2\Phi}\f{F^2}{C^2 E^4}+
 2e^{2\Phi}\f{F'^{2}}{B^2 D^2 E^2}\bigg)\bigg]+cos\theta_1cos\theta_2~R_{\psi\psi}\eea

\bea
R_{\phi_1\theta_2}&=&sin\theta_1sin\psi\bigg[\f{C^2}{4D^2}+\f{D^2}{16C^2}-\f{C^2}{4E^2}+\f{D^4}{16C^2E^2}-\f{E^2}{16C^2}-\f{E^4}{16C^2D^2}-
\f{2DA'D'}{AB^2}\nn&&+\f{DB'D'}{2B^3}-\f{DC'D'}{2B^2C}-\f{D'^2}{2B^2}+\f{2EA'E'}{AB^2}-\f{EB'E'}{2B^3}+
\f{EC'E'}{2B^2C}-\nn&&\f{DD'E'}{B^2E}
+\f{ED'E'}{B^2D}+\f{E'^2}{2B^2}-\f{DD''}{2B^2}+\f{EE''}{2B^2}
\bigg]\nn&=&-\f{1}{2}sin\theta_1sin\psi\bigg[\f{(g_s
M\alpha')^4}{64}\f{\ell^2 (E^2-D^2)}{C^2 D^4 E^4}+\f{g_s M^2
\alpha'^2}{8}e^{-\Phi}\bigg(\f{g^2_s f'^{2}}{B^2 E^2}-\f{g^2_s
k'^{2}}{ B^2D^2}+\nn&&\f{g^2_s (k-f)^{2}}{4C^2 D^2}-\f{g^2_s
(k-f)^{2}}{4C^2 E^2}-e^{2\Phi}\f{(1-F)^2}{ C^2
D^2}+e^{2\Phi}\f{F^2}{ C^2 E^2}+e^{2\Phi}\f{F'^2}{ B^2 D^2}-
e^{2\Phi}\f{F'^2}{ B^2 E^2}\bigg)\nn&&-\f{g_s M^2
\alpha'^2}{32}e^{-\Phi}(E^2-D^2)\bigg(\f{g^2_s f'^{2}}{B^2
E^4}+\f{g^2_s k'^{2}}{B^2 D^4}+\f{(k-f)^2}{2}\f{g^2_s}{C^2 D^2 E^2}+
 e^{2\Phi}\f{(1-F)^2}{C^2 D^4}\nn&&+ e^{2\Phi}\f{F^2}{C^2 E^4}+
 2e^{2\Phi}\f{F'^{2}}{B^2 D^2 E^2}\bigg)\bigg]\eea

\bea
R_{\theta_1\theta_2}&=&cos\psi\bigg[\f{C^2}{4D^2}+\f{D^2}{16C^2}-\f{C^2}{4E^2}+\f{D^4}{16C^2E^2}-\f{E^2}{16C^2}-\f{E^4}{16C^2D^2}-
\f{2DA'D'}{AB^2}\nn&&+\f{DB'D'}{2B^3}-\f{DC'D'}{2B^2C}-\f{D'^2}{2B^2}+\f{2EA'E'}{AB^2}-\f{EB'E'}{2B^3}+
\f{EC'E'}{2B^2C}-\nn&&\f{DD'E'}{B^2E}
+\f{ED'E'}{B^2D}+\f{E'^2}{2B^2}-\f{DD''}{2B^2}+\f{EE''}{2B^2}
\bigg]\nn&=&-\f{1}{2}cos\psi\bigg[\f{(g_s M\alpha')^4}{64}\f{\ell^2
(E^2-D^2)}{C^2 D^4 E^4}+\f{g_s M^2
\alpha'^2}{8}e^{-\Phi}\bigg(\f{g^2_s f'^{2}}{B^2 E^2}-\f{g^2_s
k'^{2}}{ B^2D^2}+\nn&&\f{g^2_s (k-f)^{2}}{4C^2 D^2}-\f{g^2_s
(k-f)^{2}}{4C^2 E^2}-e^{2\Phi}\f{(1-F)^2}{ C^2
D^2}+e^{2\Phi}\f{F^2}{ C^2 E^2}+e^{2\Phi}\f{F'^2}{ B^2 D^2}-
e^{2\Phi}\f{F'^2}{ B^2 E^2}\bigg)\nn&&-\f{g_s M^2
\alpha'^2}{32}e^{-\Phi}(E^2-D^2)\bigg(\f{g^2_s f'^{2}}{B^2
E^4}+\f{g^2_s k'^{2}}{B^2 D^4}+\f{(k-f)^2}{2}\f{g^2_s}{C^2 D^2 E^2}+
 e^{2\Phi}\f{(1-F)^2}{C^2 D^4}\nn&&+ e^{2\Phi}\f{F^2}{C^2 E^4}+
 2e^{2\Phi}\f{F'^{2}}{B^2 D^2 E^2}\bigg)\bigg]\eea

\bea
R_{\phi_2\theta_1}&=&sin\theta_2sin\psi\bigg[\f{C^2}{4D^2}+\f{D^2}{16C^2}-\f{C^2}{4E^2}+\f{D^4}{16C^2E^2}-\f{E^2}{16C^2}-\f{E^4}{16C^2D^2}-
\f{2DA'D'}{AB^2}\nn&&+\f{DB'D'}{2B^3}-\f{DC'D'}{2B^2C}-\f{D'^2}{2B^2}+\f{2EA'E'}{AB^2}-\f{EB'E'}{2B^3}+
\f{EC'E'}{2B^2C}-\nn&&\f{DD'E'}{B^2E}
+\f{ED'E'}{B^2D}+\f{E'^2}{2B^2}-\f{DD''}{2B^2}+\f{EE''}{2B^2}
\bigg]\nn&=&-\f{1}{2}sin\theta_2sin\psi\bigg[\f{(g_s
M\alpha')^4}{64}\f{\ell^2 (E^2-D^2)}{C^2 D^4 E^4}+\f{g_s M^2
\alpha'^2}{8}e^{-\Phi}\bigg(\f{g^2_s f'^{2}}{B^2 E^2}-\f{g^2_s
k'^{2}}{ B^2D^2}+\nn&&\f{g^2_s (k-f)^{2}}{4C^2 D^2}-\f{g^2_s
(k-f)^{2}}{4C^2 E^2}-e^{2\Phi}\f{(1-F)^2}{ C^2
D^2}+e^{2\Phi}\f{F^2}{ C^2 E^2}+e^{2\Phi}\f{F'^2}{ B^2 D^2}-
e^{2\Phi}\f{F'^2}{ B^2 E^2}\bigg)\nn&&-\f{g_s M^2
\alpha'^2}{32}e^{-\Phi}(E^2-D^2)\bigg(\f{g^2_s f'^{2}}{B^2
E^4}+\f{g^2_s k'^{2}}{B^2 D^4}+\f{(k-f)^2}{2}\f{g^2_s}{C^2 D^2 E^2}+
 e^{2\Phi}\f{(1-F)^2}{C^2 D^4}\nn&&+ e^{2\Phi}\f{F^2}{C^2 E^4}+
 2e^{2\Phi}\f{F'^{2}}{B^2 D^2 E^2}\bigg)\bigg]\eea

The Ricci scalar is \bea
R&=&\f{1}{2C^2}+\f{2}{D^2}+\f{2}{E^2}-\f{C^2}{D^2E^2}-\f{D^2}{4C^2E^2}-\f{E^2}{4C^2D^2}-\f{12A'^2}{A^2B^2}+
\f{8A'B'}{AB^3}-\nn&&\f{8A'C'}{AB^2C}+\f{2B'C'}{B^3C}-\f{16A'D'}{AB^2D}+\f{4B'D'}{B^3D}-\f{4C'D'}{B^2CD}-
\f{2D'^2}{B^2D^2}-\f{16A'E'}{AB^2E}+\nn&&\f{4B'E'}{B^3E}-\f{4C'E'}{B^2CE}-\f{8D'E'}{B^2DE}-\f{2E'^2}{B^2E^2}-
\f{8A''}{AB^2}-\f{2C''}{B^2C}-\f{4D''}{B^2D}-\f{4E''}{B^2E}
 \nn&=&\f{1}{2}\f{\Phi'^2}{B^2}+\f{g_s M^2
\alpha'^2}{16}e^{-\Phi}\bigg(\f{g^2_s f'^{2}}{B^2 E^4}+\f{g^2_s
k'^{2}}{ B^2D^4}+\f{g^2_s (k-f)^{2}}{2C^2
D^2E^2}+\nn&&e^{2\Phi}\f{(1-F)^2}{ C^2 D^4}+e^{2\Phi}\f{F^2}{ C^2
E^4}+2 e^{2\Phi}\f{F'^2}{ B^2 D^2E^2}\bigg)\eea

It just follows from the computation of the Ricci components that
all the components are not independent, which tells us to make a
choice. So we shall take the Ricci components that give independent
equations are \be R_{xx},R_{\tau\tau},R_{\theta_1\theta_1},
R_{\psi\psi}, R_{\theta_1\theta_2}.\ee

It is interesting to note that some of the Ricci components that are
computed using the metric eq(\ref{metric_ansatz}) are symmetric and
some are anti-symmetric under the interchange of
$D\longleftrightarrow E$, that is the size of the two $S^2$s.

Upon going through the ansatz to the solution, we found that there
are 9 unknowns: one from $F_3$ flux, two from $H_3$ flux, one from
dilaton and five from the metric. Simultaneously there are 9
equations: one from $F_3$ equation, two from $H_3$, one from dilaton
and five from Ricci tensor equations. So there are as many equations
as unknowns.

Let us summarize all the 9 equations:

\bea &&[{\bf 1}]~\f{1}{A^4BCD^2E^2 }\p_{\tau}[\f{A^4 C D^2 E^2
\Phi'}{B}]=-\f{g_s M^2\alpha'^{2}}{8} e^{-\Phi}[\f{g^2_s f'^{2}}{B^2
E^4}+\f{g_s k'^{2}}{B^2 D^4}+\nn&&\f{(k-f)^2}{2}\f{g^2_s}{C^2 D^2
E^2}- e^{2\Phi}\f{(1-F)^2}{C^2 D^4}-e^{2\Phi}\f{F^2}{C^2
E^4}-2e^{2\Phi}\f{F'^{2}}{B^2 D^2 E^2}],\nn &&[{\bf 2}]~\f{1-F}{2}
e^{\Phi}\f{A^4 B E^2}{CD^2}-\f{F}{2}
e^{\Phi}\f{A^4BD^2}{CE^2}+\p_\tau(F' e^{\Phi} \f{A^4C}{B})= \f{g^3_s
M^2\alpha'^2}{4} \f{A^4B}{CD^2E^2}\f{\ell(k-f)}{2}\nn&&[{\bf
3}]~\f{g_s M^2 \alpha'^2}{2}\f{A^4 B}{CD^2 E^2}\ell (1-F)=2
\p_{\tau} (f' e^{-\Phi}\f{A^4 CD^2}{BE^2})+(k-f) e^{-\Phi}\f{A^4
B}{C}\nn &&[{\bf 4}]~\f{g_s M^2 \alpha'^2}{2}\f{A^4 B}{CD^2 E^2}F
\ell=2 \p_{\tau} (k' e^{-\Phi}\f{A^4 CE^2}{BD^2})-(k-f)
e^{-\Phi}\f{A^4 B}{C},\nn&&[{\bf
5}]~\f{3A'^2}{B^2}-\f{AA'B'}{B^3}+\f{AA'C'}{B^2C}+\f{2AA'D'}{B^2D}+\f{2AA'E'}{B^2E}+\f{AA''}{B^2}=\nn&&\f{(g_s
M\alpha')^4}{64}\f{\ell^2 A^2}{C^2 D^4 E^4}+\f{g_s M^2
\alpha'^2}{32}A^2 e^{-\Phi}\bigg(\f{g^2_s f'^{2}}{B^2 E^4}+\f{g^2_s
k'^{2}}{B^2 D^4}+\nn & &\f{(k-f)^2}{2}\f{g^2_s}{C^2 D^2 E^2}+
 e^{2\Phi}\f{(1-F)^2}{C^2 D^4}+ e^{2\Phi}\f{F^2}{C^2 E^4}+
 2e^{2\Phi}\f{F'^{2}}{B^2 D^2 E^2}\bigg),\nn&&[{\bf 6}]~\f{4A'B'}{AB}+\f{B'C'}{BC}+\f{2B'D'}{BD}+\f{2B'E'}{BE}-
 \f{4A''}{A}-\f{C''}{C}-\f{2D''}{D}-\f{2E''}{E}=\nn&&\f{1}{2}\Phi'^2-\f{(g_s M\alpha')^4}{64}\f{\ell^2
B^2}{C^2 D^4 E^4}+ \f{g_s M^2 \alpha'^2}{8} e^{-\Phi}\bigg(\f{g^2_s
f'^{2}}{ E^4}+\f{g^2_s k'^{2}}{ D^4}+ 2 e^{2\Phi}\f{F'^2}{ D^2
E^2}\bigg)-\nn &&\f{g_s M^2 \alpha'^2}{32} e^{-\Phi}\bigg(\f{g^2_s
f'^{2}}{ E^4}+\f{g^2_s k'^{2}}{ D^4}+ \f{(k-f)^2}{2}\f{g^2_s
B^2}{C^2 D^2 E^2}+
 e^{2\Phi}\f{(1-F)^2B^2}{C^2 D^4}+\nn && e^{2\Phi}\f{F^2B^2}{C^2 E^4}+
 2e^{2\Phi}\f{F'^{2}}{ D^2 E^2}\bigg),\nn&&[{\bf 7}]~1-\f{C^2}{4D^2}-\f{D^2}{16C^2}-\f{C^2}{4E^2}+\f{D^4}{16C^2E^2}-
 \f{E^2}{16C^2}+\f{E^4}{16C^2D^2}-\f{2DA'D'}{AB^2}
+\nn&&\f{DB'D'}{2B^3}-\f{DC'D'}{2B^2C}-\f{D'^2}{2B^2}-\f{2EA'E'}{AB^2}+\f{EB'E'}{2B^3}-\f{EC'E'}{2B^2C}-
\f{DD'E'}{B^2E}-\nn&&\f{ED'E'}{B^2D}-
\f{E'^2}{2B^2}-\f{DD''}{2B^2}-\f{EE''}{2B^2} \nn&=&\f{(g_s
M\alpha')^4}{128}\f{\ell^2 (D^2+E^2)}{C^2 D^4 E^4}+\f{g_s M^2
\alpha'^2}{16}e^{-\Phi}\bigg(\f{g^2_s f'^{2}}{B^2 E^2}+\f{g^2_s
k'^{2}}{ B^2D^2}+\f{g^2_s (k-f)^{2}}{4C^2 D^2}+\nn&&\f{g^2_s
(k-f)^{2}}{4C^2 E^2}+e^{2\Phi}\f{(1-F)^2}{ C^2
D^2}+e^{2\Phi}\f{F^2}{ C^2 E^2}+e^{2\Phi}\f{F'^2}{ B^2 D^2}+
e^{2\Phi}\f{F'^2}{ B^2 E^2}\bigg)-\f{g_s M^2
\alpha'^2}{64}\nn&&e^{-\Phi}(D^2+E^2)\bigg(\f{g^2_s f'^{2}}{B^2
E^4}+\f{g^2_s k'^{2}}{B^2 D^4}+\f{(k-f)^2}{2}\f{g^2_s}{C^2 D^2 E^2}+
 e^{2\Phi}\f{(1-F)^2}{C^2 D^4}\nn&&+ e^{2\Phi}\f{F^2}{C^2 E^4}+
 2e^{2\Phi}\f{F'^{2}}{B^2 D^2 E^2}\bigg),\nn&&[{\bf 8}]~\f{1}{2}+\f{C^4}{D^2E^2}-\f{D^2}{4E^2}-\f{E^2}{4D^2}-
 \f{4CA'C'}{AB^2}+\f{CB'C'}{B^3}-\f{2CC'D'}{B^2D}-
\f{2CC'E'}{B^2E}-\f{CC''}{B^2}\nn &=&\f{(g_s
M\alpha')^4}{64}\f{\ell^2 }{ D^4 E^4}+\f{g_s M^2
\alpha'^2}{16}e^{-\Phi}\bigg(\f{g^2_s (k-f)^{2}}{
D^2E^2}+e^{2\Phi}\f{2(1-F)^2}{  D^4}+e^{2\Phi}\f{2F^2}{
E^4}\bigg)-\nn&&\f{g_s M^2 \alpha'^2}{32}e^{-\Phi} C^2\bigg(\f{g^2_s
f'^{2}}{B^2 E^4}+\f{g^2_s k'^{2}}{B^2
D^4}+\f{(k-f)^2}{2}\f{g^2_s}{C^2 D^2 E^2}+
 e^{2\Phi}\f{(1-F)^2}{C^2 D^4}\nn&&+ e^{2\Phi}\f{F^2}{C^2 E^4}+
 2e^{2\Phi}\f{F'^{2}}{B^2 D^2 E^2}\bigg),\nn&&[{\bf 9}]~\f{C^2}{4D^2}+\f{D^2}{16C^2}-\f{C^2}{4E^2}+\f{D^4}{16C^2E^2}-
 \f{E^2}{16C^2}-\f{E^4}{16C^2D^2}-
\f{2DA'D'}{AB^2}\nn&&+\f{DB'D'}{2B^3}-\f{DC'D'}{2B^2C}-\f{D'^2}{2B^2}+\f{2EA'E'}{AB^2}-\f{EB'E'}{2B^3}+
\f{EC'E'}{2B^2C}-\nn&&\f{DD'E'}{B^2E}
+\f{ED'E'}{B^2D}+\f{E'^2}{2B^2}-\f{DD''}{2B^2}+\f{EE''}{2B^2}
\nn&=&-\f{1}{2}\bigg[\f{(g_s M\alpha')^4}{64}\f{\ell^2
(E^2-D^2)}{C^2 D^4 E^4}+\f{g_s M^2
\alpha'^2}{8}e^{-\Phi}\bigg(\f{g^2_s f'^{2}}{B^2 E^2}-\f{g^2_s
k'^{2}}{ B^2D^2}+\nn&&\f{g^2_s (k-f)^{2}}{4C^2 D^2}-\f{g^2_s
(k-f)^{2}}{4C^2 E^2}-e^{2\Phi}\f{(1-F)^2}{ C^2
D^2}+e^{2\Phi}\f{F^2}{ C^2 E^2}+e^{2\Phi}\f{F'^2}{ B^2 D^2}-
e^{2\Phi}\f{F'^2}{ B^2 E^2}\bigg)\nn&&-\f{g_s M^2
\alpha'^2}{32}e^{-\Phi}(E^2-D^2)\bigg(\f{g^2_s f'^{2}}{B^2
E^4}+\f{g^2_s k'^{2}}{B^2 D^4}+\f{(k-f)^2}{2}\f{g^2_s}{C^2 D^2 E^2}+
 e^{2\Phi}\f{(1-F)^2}{C^2 D^4}\nn&&+ e^{2\Phi}\f{F^2}{C^2 E^4}+
 2e^{2\Phi}\f{F'^{2}}{B^2 D^2 E^2}\bigg)\bigg]\eea
\section{Solutions}
There exists three  solutions to these equations and are known as
 KT, KS and ABY/DKM.  These solutions when expressed in our parametrization,  reads as\\
\underline{KT:}\\
\bea &&A=B^{-1}=h^{-\frac{1}{4}},~~\tau=r,~~C^2=h^{\frac{1}{2}}
\frac{r^2}{9},~~D^2=E^2=h^{\frac{1}{2}} \frac{r^2}{6},\nn&&
\Phi=Log~g_s,F=\f{1}{2},~~f=k=\f{3}{2}Log~\f{r}{r_0},~~\ell=\f{3}{2}Log\f{r}{r_0}\nn&&
h(r)=\frac{27 \pi g_s \alpha'^2}{4r^4}[\frac{3g_s
M^2}{2\pi} Log ~\frac{r}{r_0}+\frac{3g_s M^2}{8\pi}] \eea\\

\underline{KS:}\\
\bea
&&A^2=h^{-\f{1}{2}},~~B^2=C^2=\f{h^{\f{1}{2}}\varepsilon^{\f{4}{3}}}{6
K^2},~~D^2=\f{h^{\f{1}{2}}\varepsilon^{\f{4}{3}}}{2} K
cosh^2\f{\tau}{2},~~E^2=\f{h^{\f{1}{2}}\varepsilon^{\f{4}{3}}}{2} K
sinh^2\f{\tau}{2},\nn&&
K=\f{(sinh2\tau-2\tau)^{\f{1}{3}}}{2^{\f{1}{3}}sinh\tau},\Phi=Log~g_s,
F=\f{sinh\tau-\tau}{2 sinh\tau},~~f=\f{\tau coth\tau-1}{2
sinh\tau}(cosh\tau-1),\nn&&k=\f{\tau coth\tau-1}{2
sinh\tau}(cosh\tau+1),~~\ell=\f{\tau coth\tau-1}{4
sinh^2\tau}(sinh2\tau-2\tau),\nn&& h(\tau)=(g_s M \alpha')^2
2^{\f{2}{3}} \varepsilon^{-\f{8}{3}} I(\tau),~~
I(\tau)=\int^{\infty}_{\tau} dx \f{x
coth~x-1}{sinh^2x}(sinh2x-2x)^{\f{1}{3}},\nn&&I(\tau)\approx
0.71805+0.18344~\tau^2-0.0306 ~\tau^4+\cdots\eea

\underline{ABY/DKM:}\\
 The parametrization that we used is related
to  DKM as  \bea &&A^2=B^{-2}=r^2
e^{2a},~~C^2=\f{e^{2b-2a}}{9},~~D^2=E^2=\f{e^{2c-2a}}{6},~~\tau=r\nn&&F=\f{1}{2},~~3
(g_s M \alpha') f=3 (g_s M \alpha') k=k_{DKM}\eea

 If we assume that the unknown functions that appear in $B_2$ and
 metric satisfies the following conditions i.e.
 $f=k$ and $D^2=E^2$ then there is a simple
solution to  eq.(\ref{H_3_eq}) that is $F=\f{1}{2}$, which is
consistent with the flux quantization condition for $F_3$ and is in
fact the solution for KT and ABY/DKM solutions. For this special
case we left with 6 unknowns and as many equations.

Let us try to find the linearized solution to the rest of the
equations of motion written above, for this special case. Upon
assuming that the solution depends on two parameters ${\cal S}$ and
$ \phi$, explicitly it means that the solution reads as \bea
&&h(r)=\frac{27 \pi g_s \alpha'^2}{4r^4}[\frac{3g_s M^2}{2\pi} Log
~\frac{r}{r_0}+\frac{3g_s M^2}{8\pi}]+\nn&&\f{(g_S M
\alpha')^2}{r^8}[(h_1+h_2 Log~r){\cal S}-h_3 \phi]\nn&&
A(r)=h^{-\f{1}{4}}=B^{-1},~~C(r)=\f{r}{3}h(r)^{\f{1}{4}}[1+\f{1}{r^4}\bigg((c_1+c_2
Log~r){\cal S}-c_3 \phi\bigg)]\nn&&
D(r)=\f{r}{\sqrt{6}}h(r)^{\f{1}{4}}[1+\f{1}{r^4}\bigg((d_1+d_2
Log~r){\cal S}-d_3 \phi\bigg)]=E(r)\nn&&
\Phi(r)=Log~g_s+\f{1}{r^4}\bigg((p_1+p_2 Log~r){\cal S}-p_3
\phi\bigg)\nn && f(r)=k(r)=\ell(r)=\f{3}{2}
Log~r+\f{1}{r^4}\bigg((f_1+f_2 Log~r){\cal S}-f_3 \phi\bigg),\eea

where
$c_1,~c_2,~c_3,~d_1,~d_2,~d_3,~f_1,~f_2,~f_3,~h_1,~h_2,~h_3,~p_1,~p_2,p_3
$ are all constants. Now solving these 6 equations we get the answer
which depends on $h_1,~h_2$ and $h_3$ as
\bea\label{gen_fluctuation_sol}
&&c_1=\f{8}{81}h_2,~~c_2=0,~~c_3=0,~~d_1=0,~~d_2=0,~~d_3=0,,\nn&&
~p_1=-\f{64}{81}h_1+\f{52}{81}h_2,~~p_2=-\f{16}{27}h_2,~~p_3=-\f{64}{81}h_3,\nn&&f_1=\f{8}{27}h_1-\f{h_2}{27},~~
f_2=\f{4}{9}h_2,~~f_3=\f{8}{27}h_3,\eea

it essentially means that this  linearized perturbation generates
solution which is characterized  completely by the warp factor.

 For a choice like: $h_1=\f{1053}{256},~~h_2=\f{81}{16}$, and
$h_3=\f{81}{64}$ , we get back the DKM solution \cite{dkm}. Now, the
question arises why such a choice is special ?

To answer this question we may need to look at the supersymmetry
preserved by the solution. This we can say by looking at the type of
the complex combination of three form flux that comes from RR, $F_3$
and NS-NS sector, $H_3$.

The general form of the complex 3-form \bea G_3&=&F_3-i
e^{-\Phi}H_3\nn&=&\frac{M \alpha'}{2}[(1-F)g_5\w g_3\w g_4+Fg_5\w
g_1\w g_2-i e^{-\Phi} g_s \frac{(k-f)}{2} g_5\w g_1\w g_3-\nn&& i
e^{-\Phi} g_s \frac{(k-f)}{2} g_5\w g_2\w g_4+F' d\tau\w g_1\w g_3
+F' d\tau\w g_2\w g_4-\nn&& i e^{-\Phi} g_s f'd\tau\w g_1\w g_2-i
e^{-\Phi} g_s k'd\tau\w g_3\w g_4]\eea

For KT case it reduces to \be G_3=\frac{M
\alpha'}{4}(g_5-2ie^{-\Phi} g_s r f' \frac{dr}{r})\w(g_1\w g_2+g_3\w
g_4)\ee

Using the solution eq(\ref{gen_fluctuation_sol}), the metric for KT
case reads \be\label{changed_conifold_metric}
ds^2=h^{-\frac{1}{2}}\eta_{\mu\nu}dx^{\mu}dx^{\nu}+r^2
h^{\frac{1}{2}}[\frac{dr^2}{r^2}+(1+2\frac{c_1 {\cal
S}}{r^4})\frac{g^2_5}{9}+\frac{g^2_1+g^2_2+g^2_3+g^2_4}{6}]\ee

From this metric it just follows that we can introduce complex
coordinate as \be \frac{dr}{r}+i (1+\frac{c_1 {\cal S}}{r^4})
\frac{g_5}{3}=\frac{2}{3r^3} \overline{z}_idz_i,~~\frac{dr}{r}-i
(1+\frac{c_1 {\cal S}}{r^4}) \frac{g_5}{3}=\frac{2}{3r^3}
{z_i}\overline{dz}_i,\ee and from the paper of \cite{hko} \be g_1\w
g_2+g_3\w g_4=\frac{2i}{r^6} \epsilon_{ijkl} z_i
\overline{z}_jdz_k\w \overline{dz}_l.\ee

combining all that we get the expression \bea G_3&=&\frac{M
\alpha'}{2 r^9} \bigg[\bigg((1-\frac{c_1 {\cal S}}{r^4})+\frac{2}{3}
e^{-\Phi}g_s r f'\bigg)\overline{z}_m dz_m\w\epsilon_{ijkl}z_i
\overline{z}_jdz_k\w \overline{dz}_l -\nn&&\bigg((1-\frac{c_1 {\cal
S}}{r^4})-\frac{2}{3} e^{-\Phi}g_s r f'\bigg)z_m
\overline{dz}_m\w\epsilon_{ijkl}z_i \overline{z}_jdz_k\w
\overline{dz}_l\bigg]\nn&=&\frac{1}{2i}(G_++G_-)\eea

It just follows trivially that \bea G_+&=&\frac{i M
\alpha'}{r^9}\bigg((1-\frac{c_1 {\cal S}}{r^4})+\frac{2}{3}
e^{-\Phi}g_s r f'\bigg)\nn&=&\frac{i M
\alpha'}{r^9}\bigg[2-\frac{{\cal S}
h_2}{r^4}\bigg(\frac{28}{81}+\frac{16}{27}
Log~r\bigg)\bigg]\overline{z}_m dz_m\w\epsilon_{ijkl}z_i
\overline{z}_jdz_k\w \overline{dz}_l\nn G_-&=&-\frac{i M
\alpha'}{r^9}\bigg((1-\frac{c_1 {\cal S}}{r^4})-\frac{2}{3}
e^{-\Phi}g_s r f'\bigg)\nn&=&-\frac{i M
\alpha'}{r^9}\bigg[\frac{{\cal S}
h_2}{r^4}\bigg(\frac{12}{81}+\frac{16}{27} Log~r\bigg)\bigg]z_m
\overline{dz}_m\w\epsilon_{ijkl}z_i \overline{z}_jdz_k\w
\overline{dz}_l,\nn\eea where $G_+$ is a (2, 1) form and $G_-$ is
(1, 2) form and its computed using the back reacted metric.
 Now it is easy to draw the conclusion that for $h_2=0$ we can
have a supersymmetry preserving solution. Which is not surprising,
as the metric of the conifold eq(\ref{changed_conifold_metric}) has
changed to terms proportional to $c_1\sim h_2$.

Probably it makes sense to say that we have a two dimensional real
space that is described by $(h_1,h_3 )$ plus a point at the origin
$h_2=0$, preserves supersymmetry. As soon as we go away from the
origin along the $h_2$ axis the system is not any more
supersymmetric. Hence the interpretation that we have a
supersymmetry preserving plane is correct.

After taking the back reaction of the fluctuation we see that the
metric of changed singular metric can be made to be same as the
metric of the singular conifold metric upon setting the
non-supersymmetric fluctuation to zero. This point could be useful
to find the solution of the linearized fluctuation of the deformed
conifold \cite{ssp}.

\section{conclusion}

We have re-visited the linearized perturbation \cite{dkm} to the
gravity solution  of the intersecting D3 branes and D5 branes
wrapped on a 2 sphere \cite{kt}, linear  in the parameters ${\cal
S}$ and $\phi$ and found that there arises infinitely many choices
to $h_1,~~h_2$ and $h_3$.  For any choice except the choice
$(h_1,~~h_3)$ plane sitting at $h_2=0$,  break supersymmetry
dynamically.

For a supersymmetry preserving solution the vacuum energy should
vanish, which means from the dual field theory point of view the
energy should vanish and by Lorentz invariance the energy-momentum
tensor should go as \be < T_{\mu\nu}> ~~\sim ~~ \eta_{\mu\nu}~h_2
{\cal S}.\ee

Its exact structure need to be computed following \cite{aby}, but
for the  specific choice to $h_1,~h_2$ and $h_3$ i.e. the DKM
solution \cite{dkm}, the authors have given the relation between the
parameter ${\cal S}$ and the energy momentum tensor, $T_{\mu\nu}$.

Let us recall from the gauge gravity duality for the cascading
theory, that is the two gauge couplings are related to the bulk
fields \bea
&&\frac{4\pi^2}{g^2_1}+\frac{4\pi^2}{g^2_2}=\frac{\pi}{g_s}
e^{-\Phi}\nn&&
\bigg(\frac{4\pi^2}{g^2_1}-\frac{4\pi^2}{g^2_2}\bigg)g_s
e^{\Phi}=\frac{1}{2\pi \alpha'}\bigg(\oint_{S^2}B_2\bigg)-\pi ~(mod
~2\pi)\eea

Now using the solution to the bulk field equations of motion into
this duality \bea
&&\frac{4\pi^2}{g^2_1}+\frac{4\pi^2}{g^2_2}=\frac{\pi}{g^2_s}\bigg[
1-\f{1}{r^4} \bigg((-\frac{64}{81} h_1+\frac{52}{81}h_2){\cal
S}-\frac{16}{27}h_2 {\cal S} Log~r+\frac{64}{81}h_3 \phi
\bigg)\bigg]\nn &&
\frac{4\pi^2}{g^2_1}-\frac{4\pi^2}{g^2_2}=\frac{3M}{g_s}
Log~r+\nn&&\frac{{\cal S}}{r^4}
(-\frac{64}{81}h_1+\frac{52}{81}h_2-\frac{16}{27}h_2
Log~r)(\pi-\frac{3M}{g_s} Log~r)+\nn&&\frac{{\cal
S}}{r^4}\frac{2M}{g_s}\bigg((\frac{8}{27}h_1-\frac{h_2}{27})+\frac{4}{9}h_2Log~r\bigg)-\nn&&\frac{\phi}{r^4}\bigg(
-\frac{64}{81}h_3 (\pi-\frac{3M}{g_s} Log~r)+\frac{2M}{g_s}
\frac{8}{27}h_3\bigg)-\pi ~(mod ~2\pi)\eea

The non constancy nature of the dilaton makes that the $\beta$
function for $\frac{8\pi^2}{g^2_1}+\frac{8\pi^2}{g^2_2}$ do not
vanishes any more and $\frac{8\pi^2}{g^2_1}-\frac{8\pi^2}{g^2_2}$
has the leading term that goes logarithmically and the subleading
term goes as inverse power law.

Its important to understand the field theory dual of this solution
and the connection of it with \cite{iss}-\cite{frv}, if any.

{\bf Acknowledgement}\\
I would like to thank  Shamit Kachru for useful correspondences,
 Dipankar Das, Srijesh and the computer center, IOPB for computer related help.
 I am  greatly helped by
Hara Lenka and A. Rath. It gives me immense pleasure to thank my
family for their love, affection and support. Also I have enjoyed
the company of my fellow villagers, Barchana.


\begin{thebibliography}{99}
\bibitem{jm} J. Maldacena,  ``The large N limit of superconformal field theories
and supergravity," Adv. Theor. Math. Phys. 2, 231 (1998), Int. J.
Theor. Phys. 38, 1113, (1999), hep-th/9711200.
\bibitem{gkp1} S. Gubser, I. Klebanov and A. Polyakov ``Gauge theory
correlators from non-critical string theory," Phys. Lett. B 428, 105
(1998), hep-th/9802109
\bibitem{ew}E. Witten ``Anti-de Sitter space and holography," Adv.
Theor. Math. Phys. 2, 253 (1998), hep-th/9802150.
\bibitem{kn} I. Klebanov and N. Nekrasov  ``Gravity duals of fractional branes and
logarithmic RG flow," Nucl. Phys. B 574, 263 (2000), hep-th/9911096.
\bibitem{kt} I. Klebanov and A. Tseytlin ``Gravity duals of supersymmetric SU(N)x SU(N+M)
gauge theories," Nucl. Phys. B 578, 123 (2000), hep-th/0002159.
\bibitem{mt} R. Minasian and D. Tsimpis ``On the geometry of non-trivially embedded branes,''hep-th/9911042.
\bibitem{ks} I. Klebanov and M. Strassler ``Supergravity and a confining gauge theory: duality
cascades and $\chi$SB-resolution of naked singularity," JHEP 0008,
052 (2000), hep-th/0007191.
\bibitem{cd} P. Candelas and X. de la Ossa ``Comments on conifolds,"
Nucl. Phys. B 342, 246 (1990).
\bibitem{ns} N. Seiberg ``Electric-magnetic duality in supersymmetric non-Abelian gauge
theories," Nucl. Phys. B 435, 129 (1995), hep-th/9411149.
\bibitem{ms} M. Strassler ``The duality cascade," hep-th/0505153.
\bibitem{gkp} S. Giddings, S. Kachru and J. Polchinski ``Hierarchies from fluxes in string
compactifications," Phys. Rev. D 66, 106006 (2002), hep-th/0105097.
\bibitem{kow} I. Klebanov, P. Ouyang and E. Witten ``A gravity dual
of the chiral anomaly," Phys. Rev. D 65, 105007 (2002),
hep-th/0202056.
\bibitem{au} A. Uranga  ``Brane configurations for branes at
conifolds," JHEP 01, 022 (1999)
\bibitem{dm} K. Dasgupta and S. Mukhi ``Brane constructions,
conifolds and M-theory," Nucl. Phys. B 551, 204 (1999),
hep-th/9811139.
\bibitem{pt} G. Papadopoulos and A. Tseytlin ``Complex geometry of
conifolds and 5-brane wrapped on 2-sphere," Class. Quant. Grav. 18
1333 (2001), hep-th/0012034.
\bibitem{dkm} O. DeWolfe, S. Kachru and M. Mulligan ``A gravity dual
of metastable dynamical supersymmetry breaking," arXiv:0801.1520
\bibitem{hko} C. Herzog, I. Klebanov and P. Ouyang ``Remarks on the
warped deformed conifold," hep-th/0108101.
\bibitem{aby} O. Aharony, A. Buchel and A. Yarom ``Holographic
renormalization of cascading gauge theories," Phys. Rev. D 72,
066003 (2005), hep-th/0506002.
\bibitem{ssp} \textsl{Work in progess}.
\bibitem{iss} K. Intriligator, N. Seiberg and D. Shih ``Dynamical
SUSY breaking in meta-stable vacua," JHEP 0604, 021 (2006),
hep-th/0602239.
\bibitem{dst} M. Douglas, J. Shelton and G. Torroba ``Warping and supersymmetry breaking,"
arXiv:0704.4001.
\bibitem{it} K. Intriligator and S. Thomas ``Dual descriptions of
supersymmetry breaking," hep-th/9608046.
\bibitem{clp} Z. chacko, M. Luty and E. Ponton ``Calculable dynamical
supersymmetry breaking on the deformed moduli spaces," JHEP 9812,
016 (1998), hep-th/9810253.
\bibitem{pt} E. Poppitz and S. Trivedi ``Dynamical supersymmetry
breaking," Ann. Rev. Nucl. Part. Sci. 48, 307 (1998),
hep-th/9803107.
\bibitem{iy} K. Izawa and T. Yanagida ``Dynamical supersymmetry
breaking in vector-like gauge theories," Prog. Theor. Phys. 05 829
(1996), hrp-th/9602180.
\bibitem{js} S. Kuperstein and J. Sonnenschein ``Analytic
nonsupersymmetric backgrond dual to a confining gauge theory and the
corresponding plane wave theory of hadrons," JHEP 0402, 015 (2004),
hep-th/0309011.
\bibitem{abfs1} R. Argurio, M. Bertolini, S. Franco and S. Kachru
``Gauge/gravity duality and metastable dynamical supersymmetry
breaking," JHEP 0701, 083 (2007), hep-th/0610212
\bibitem{abfs2}R. Argurio, M. Bertolini, S. Franco and S. Kachru
``Metastable vacua and D-branes at the conifold," JHEP 0706, 017
(2007), hep-th/0703236.
\bibitem{absv} M. Aganagic, C. Beem, J. Seo and C. Vafa
``Geometrically induced metastability and holography," Nucl. Phys. B
789, 382 (2008), hep-th/0610249.
\bibitem{hsv} J. Heckman, J. Seo and C. Vafa ``Phase structure of
brane/anti-brane system at large N," JHEP 0707, 073 (2007),
hep-th/0702077.
\bibitem{mps} J. Marsano, K. Papadodimas and M. shigemori
``Nonsupersymmetric brane/antibrane configurations in type IIA and M
theory," arXiv:0705.0983.
\bibitem{aks} O. Aharony, S. Kachru and E. Silverstein ``Simple
stringy dynamical susy breaking," arXiv:0708.0493.
\bibitem{abf} M. Aganagic, C. Beem and B. Freivogel ``Geometric
metastability, quivers and holography," arXiv:0708.0596
\bibitem{abs} M. Aganagic, C. Beem and S. Kachru ``Geometric
transitions and dynamical susy breaking," arXiv:0709.4277
\bibitem{bmv} M. Buican, D. Malyshev and H. Verlinde ``On the
geometry of metastable supersymmetry breaking," arXiv:0710.5519.
\bibitem{stud}G. Shiu, G. Torroba, B. Underwood and M. Douglas ``Dynamics of warped flux compactification,''arXiv:0803.3068.
\bibitem{frv}S. Franco, D. Rodriguez-Gomez and H. Verlinde ``N-ification of forces: A holographic perspective on D-brane model building,''arXiv:0804.1125.

\end{thebibliography}
\end{document}